\documentclass[12pt]{article}
\usepackage{amsmath,amssymb}
\usepackage{graphicx,color}
\usepackage{cite}
\usepackage{bm}
\usepackage{dcolumn}
\newcommand{\be}{\begin{equation}}
\newcommand{\bea}{\begin{eqnarray}}
\newcommand{\eea}{\end{eqnarray}}
\newcommand{\ba}{\begin{array}}
\newcommand{\ea}{\end{array}}
\newcommand{\ee}{\end{equation}}

\expandafter\ifx\csname mathbbm\endcsname\relax

\else

\fi \textheight 22cm \textwidth 15cm \topmargin 1mm
\def\a{\alpha}
\begin{document}

\begin{titlepage}
\hfill
\vbox{
    \halign{#\hfil         \cr
           hep-th/0602022 \cr
	   IP/BBSR/2006-9 \cr
           IPM/P-2006/008 \cr
           TIFR/TH/06-04  \cr
           IC/2006/005  \cr
           } 
      }  
\vspace*{10mm}
\begin{center}
{\Large {\bf Non-Supersymmetric Attractors in $R^2$ Gravities}\\ }
\vspace*{15mm} \vspace*{1mm} {B. Chandrasekhar$^{a,b}$
\footnote{chandra@iopb.res.in, $a$: Present address}, S. Parvizi$^b$ }
\footnote{parvizi@theory.ipm.ac.ir}, A. Tavanfar$^{b,c}$
\footnote{art@ipm.ir} and H. Yavartanoo$^{b,d}$
\footnote{myavarta@ictp.it}
\\
\vspace*{1cm}
{$^a$ Institute of Physics, Bhubaneswar 751 005, India 
\\ \vspace{3mm}
$^b$ Institute for Studies in Theoretical Physics
and Mathematics (IPM)\\
P.O. Box 19395-5531, Tehran, Iran \\
\vspace*{3mm}
$^c$Tata Institute of Fundamental Research \\
Homi Bhabha Road, Mumbai, 400 005, INDIA\\
\vspace*{3mm}
$^d$High Energy Section, \\
The Abdus Salam International Centre for Theoretical Physics,
\\Strada Costiera, 11-34014 Trieste, Italy.\\
}
\end{center}
\begin{abstract}
We investigate the attractor mechanism for spherically symmetric
extremal black holes in a theory of general $R^2$ gravity in
$4$-dimensions, coupled to gauge fields and moduli fields. 
For the general $R^2$ theory, we look for solutions which are 
analytic near the horizon, show that they exist and enjoy the attractor behavior. 
The attractor point is determined by extremization of an effective potential 
at the horizon. This analysis includes the backreaction and supports the validity of 
non-supersymmetric attractors in the presence of
higher derivative interactions. To include a wider class of solutions, we continue 
our analysis for the specific case of a Gauss-Bonnet theory which is non-topological, 
due to the coupling of Gauss-Bonnet terms to the moduli fields. We find that 
the regularity of moduli fields at the horizon is sufficient for 
attractor behavior. For the non-analytic sector, this regularity condition in turns implies the minimality of the effective potential at the attractor point.
\end{abstract}
\end{titlepage}
\section{Introduction} 
Supersymmetric BPS black holes in string theory are known to exhibit an attractor 
mechanism, whereby the values of scalar fields at the horizon are determined only 
in terms of the charges carried by the black hole and are independent of the 
asymptotic values of these scalar fields \cite{{Ferrara:1995ih},{Ferrara:1996dd},
{Ferrara:1996um}}. As a result, the black hole solution at the horizon  and the 
resulting entropy turn out to be determined completely in terms of the conserved 
charges associated with the gauge fields. Following this, the attractor mechanism has 
been studied extensively in the context of the supergravity and string theory. 

\vskip 0.2cm

It was further discussed in \cite{{Ferrara:1997tw},{Gibbons:1996af}},
that the attractor mechanism can also work for non-supersymmetric cases. Recently 
there has been a surge  of interest in studying the attractor mechanism without 
the use of supersymmetry. In particular, important developments are taking place 
in the context of non-supersymmetric attractor mechanism, in the presence of higher 
curvature interactions in theories with and without supersymmetry.
\vskip 0.2cm

The role of non-supersymmetric attractors was further clarified in \cite{Goldstein:2005hq}. By using perturbative methods and numerical analysis it was shown that the attractor mechanism can work for non-supersymmetric extremal black holes. The authors considered theories with gravity, gauge fields and scalars in four and higher dimensions which are asymptotically flat or Anti-De Sitter. It is worth mentioning that the scalar fields do not contain any potential term in the action, so as to allow them to behave as moduli at infinity. However, the coupling of these scalars with the gauge fields acts like an effective potential for the scalars. This is with the foresight to fix the scalar field values in terms of the conserved charges associated with the gauge fields. Varying the effective 
potential, the values of scalar fields are determined in terms of the charges carried by the black hole. Several examples were presented in \cite{Goldstein:2005hq} and also in the context of string theory \cite{Tripathy:2005qp}. Recently in \cite{Goldstein:2005rr}, a c-function was introduced which monotonically decreases from infinity to the horizon and coincides with the area of the horizon at the horizon. Further developments are studied in \cite{{Kallosh:2005bj},{Kallosh:2005ax},{Cardoso:2006bg}}.

\vskip 0.2cm

Following these developments, it is important to understand whether the attractor mechanism can work in the absence of supersymmetry when there are higher derivative terms in the action. In \cite{Sen:2005wa}, using the near horizon geometry of extremal black hole to be $AdS_2 \times S^2$, it was explicitly shown that extremizing the `entropy function' with respect to the scalar fields, leads to a generalized attractor mechanism 
(see also~\cite{Kraus:2005vz}). The Legendre transform  of this function with respect to the electric charges, evaluated at the extermum, is proportional to the entropy of the black hole. This method has been used to compute corrections to the entropy of different black holes due to the higher corrections to the effective action \cite{{Sen:2005iz},{Prester:2005qs},{Sinha:2006yy},{Alishahiha:2006ke}}.
In fact, the addition of a Gauss-Bonnet term reproduces the entropy associated with four charge black holes in the Heterotic string theory, which were earlier computed making heavy use of supersymmetry. All these results also agree with the microscopic counting of entropy \cite{{9711053},{9602060},{9603191},{9801081},{9812082},{9904005},{9906094},{9910179},{0007195},{0009234},{0012232}}. 
However the existence of a solution is a presumption in \cite{Sen:2005wa}, 
though it has the advantage that the action is very general.
\vskip 0.2cm
Recently it has been shown that the Legendre transformation of  black hole entropy with respect to the electric charges is related to the generalized prepotential, and this led to the new conjectured relation between the black hole  entropy and topological string partition function \cite{{0405146},{0412139},{0502211},{0504221}}. Applying the results of these black holes to special case of black holes in Heterotic string theory with purely electric charges, one finds agreement between black hole entropy and the degeneracy of elementary string states \cite{{0409148},{0410076},{0411255},{0411272},{0501014},{0502126},{0502157},{0504005},{0505122}} even for small black holes with vanishing classical horizon area \cite{{9504147},{9506200},{9712150}}.
\vskip 0.2cm
Following the approach in \cite{Goldstein:2005hq}, we show that the attractor mechanism works for non-supersymmetric theories even after the inclusion of higher derivative $R^2$ terms, and in a more specific case, the Gauss-Bonnet terms. In  four dimensions, since the Gauss-Bonnet term is a total derivative, its coupling with dilaton field plays an important role. Without this coupling, Gauss-Bonnet term dose not contribute to the equations of motion. Thus, after coupling to the scalar fields we find non-trivial equations of motions and solve all of them together. 
\vskip 0.2cm
To solve the equations, we consider the background to be an asymptotically flat extremal black hole and look for scalar fields which are free at infinity and regular at the horizon. It will be shown that subject to these boundary conditions, all solutions of the scalar fields get fixed values at the horizon, independent of their arbitrary moduli values in the asymptotic region. This analysis supports 
the existence of attractor mechanism for black holes in higher derivative gravity. 
\vskip 0.2cm
Since we are faced the situation of dealing with a set of coupled non-linear equations involving 
the metric and many
scalar fields, we solve them by a series solution method. In $R^2$ gravity we derive the solutions as a series with analytic behavior near the horizon. For completeness, we consider non-analytic, but regular solutions in the Gauss-Bonnet case.  We also make use of 
a perturbative method in terms of the parameter $\a'$, for the case of Gauss-Bonnet theory and construct the perturbative solutions on top of the zeroth order solutions of the Einstein-Hilbert gravity given in \cite{Goldstein:2005hq}. In all these cases, we show 
that, as long as regularity and asymptotic flatness conditions are satisfied, the solutions enjoy the attractor mechanism.
\vskip 0.2cm
The organization of the paper is as follows. In the next section we give a brief review of the non-supersymmetric attractor mechanism in the Einstein-Hilbert theory. In section 3, we discuss the attractor mechanism near the horizon of an extremal black hole,
for general higher derivative theories. Section 4 is devoted to $R^2$ gravity where we find an analytic series solution for the set of coupled equations. In section 5, we investigate Gauss-Bonnet theory and give as large as possible class of solutions with given boundary conditions. We conclude in section 6.

\section{Brief Review of Non-Supersymmetric Attractor Mechanism} 
Here we collect some important points regarding the non-supersymmetric attractor mechanism in four dimensional asymptotically flat space-time for later use. It is instructive to start with gravity theories coupled to $U(1)$ gauge fields and scalar fields, dictated by following the bosonic action\footnote{In the convention of \cite{Goldstein:2005hq}, $\frac{1}{\sqrt{-g}}$ factor is involved in the definition of $\epsilon_{\mu\nu\rho\sigma}$.}:
\be
\label{EHaction}
  S=\frac{1}{\kappa^{2}}\int d^{4}x\sqrt{-g}\left(R-2(\partial\phi_i)^{2}-
  f_{ab}(\phi_i)F^a_{\mu \nu} F^{b \ \mu \nu} - \frac{1}{2\sqrt{-g}}{\tilde f}_{ab}(\phi_i) F^a_{\mu \nu}
  F^b_{\rho \sigma} \epsilon^{\mu \nu \rho \sigma} \right) 
 \ee
where $F^a_{\mu\nu}, \; a=0,...N$ are gauge fields and $\phi^i, \; i=1,...n$ are scalar fields. The scalars have no potential terms but determine the gauge coupling constant. We note that $g_{ij}$ refers to the metric in the moduli space and this is different from space time metric, i.e., $g_{\mu\nu}$.
The equations of motion derived from the action (\ref{EHaction}) are as follows:
\bea
&&R_{\mu\nu}-2\partial_{\mu}\phi_i\partial_{\nu}\phi_i
  =   f_{ab}(\phi_i) \left(2F^a_{\phantom{a}\mu\lambda}
F^{b\phantom{\nu}\lambda}_{\phantom{b}\nu}-
    \frac{1}{2}g_{\mu\nu}F^a_{\phantom{a}\kappa\lambda}
F^{b \kappa\lambda} \right),
\cr\cr
&&\frac{1}{\sqrt{-g}}\partial_{\mu}(\sqrt{-g}\partial^{\mu}\phi_i)
= \frac{1}{4} \partial_i(f_{ab}) F^a _{\phantom{a}\mu\nu}
F^{b \mu\nu} -\frac{1}{8} \frac{1}{\sqrt{-g}}\partial_i({\tilde f}_{ab}) 
F^a_{\mu \nu}  F^b_{\rho \sigma} \epsilon^{\mu \nu \rho \sigma}, \cr\cr
&&\partial_{\mu}\left(\sqrt{-g} f_{ab}(\phi_i) F^{b \mu\nu}  +
\frac{1}{2} {\tilde f}_{ab}
  F^b_{\rho \sigma} \epsilon^{\mu \nu \rho \sigma}
 \right) = 0.
\eea
A spherically symmetric space-time metric in 3+1 dimensions can be taken to be of
the form:
\bea
\label{metric}
ds^2&=&-a(r)^2dt^2+a(r)^{-2}dr^{2}+b(r)^2d\Omega^2 
\eea
On the other hand, the Bianchi identity and equations of motion of gauge fields can be solved by taking the gauge field strengths to be of the form:
\be \label{gaugefield}
F^a=f^{ab}(\phi_i)(Q_{eb}-\tilde{f}_{bc}Q^c_m)\frac{1}{b^2}dt\wedge dr + Q^a_m \sin\theta d\theta \wedge d\varphi,
\ee
where $Q^a_m$ and $Q_{ea}$ are constants that determine the magnetic and electric charges carried by the gauge fields $F^a$, and $f^{ab}$ is inverse of $f_{ab}$.
Using the equations of motion it is possible to derive the following second order equations for the unknown functions $a(r)$, $b(r)$ \cite{Goldstein:2005hq}:
\bea
 (a^{2}(r)b^{2}(r))^{''} &=& 2, \\ \label{eqa1}
\frac{b''}{b} &=& -(\partial_r \phi)^2, \\ \label{eqa2}
\partial_r(2a^2b^2\partial_r\phi_i) &=& \frac{\partial_i V_{eff}}{b^2}, \label{eqphi} \\
-1+a^2b'^2+\frac{a^{2'}b^{2'}}{2} &=& -\frac{1}{b^2} (V_{eff}(\phi_i)) +a^2b^2(\partial_r\phi)^2,
\label{const1}
\eea
with the effective potential given by,
\be \label{eff}
V_{eff}(\phi_i) = f^{ab}(Q_{ea}-\tilde{f}_{ac}Q^c_{m})(Q_{eb}-\tilde{f}_{bd}Q^d_{m}) + f_{ab}Q^a_{m}Q^b_m.
\ee
It is useful to note that the equations of motion (\ref{eqa1}-\ref{eqphi}) given above,
can be derived from the following one-dimensional action:
\bea \label{11daction}
  S &=& \frac{1}{\kappa^{2}}\int dr\left[2-(a^{2}b^{2})^{''}-2a^{2}bb^{''}
    -2a^{2}b^{2}(\partial_{r}\phi)^{2}
-\frac{2 V_{eff}(\phi_i)}{b^{2}} \right].
\eea
Furthermore, eq. (\ref{const1}) stands for the Hamiltonian constraint which must 
be imposed in addition. 
\vskip 0.2cm
Now two sufficient conditions having the attractor behavior for the moduli fields 
can be stated as follows \cite{Goldstein:2005hq}. First, for fixed charges, 
$V_{eff}$ as a function of the moduli, must have a critical point. Let us denote 
the critical values of the scalars as $\phi^i=\phi_0^i$. Then we have,
\be
\partial_i V_{eff}(\phi_{i0})=0.
\ee
Second, the matrix of second derivatives of the potential at the critical point,
\be
\label{matrix}
M_{ij}=\frac{1}{2} \partial_i\partial_j V_{eff}(\phi_{i0}),
\ee
should have positive eigenvalues. Schematically one can write,
\be
M_{ij} > 0.
\ee
This condition guarantees the stability of the solution. Let us 
refer to $M_{ij}$ as the mass matrix and the corresponding eigenvalues 
as masses (more correctly ${\rm mass}^2$ terms) for the fields, 
$\phi$ \cite{Goldstein:2005hq}.
\vskip 0.2cm
Once the two conditions mentioned above are met, it was argued in \cite{Goldstein:2005hq} that the attractor mechanism typically works. There is an extremal Reissner-Nordstrom black hole solution in the theory, where the black hole carries the charges specified by the parameters, $Q_m^a, Q_{ea}$. Further, the moduli take critical values, $\phi_0$ at 
the horizon, which are independent of their values at infinity. In other words, 
although $\phi$ is free at infinity as a moduli field, it is attracted to a fixed 
point, $\phi_0$ at the horizon.
\vskip 0.2cm

Let us now consider a constant $\phi_0$ solution, in which case, 
the resulting horizon radius can be shown to be,
\be
b_H^2 = V_{eff}(\phi_{i0}),
\ee
with the entropy given as:
\be
S_{BH}=\frac{1}{4}A=\pi b_H^2
\ee
In the next section, we study the behavior of solutions near 
the horizon and discuss the attractor conditions for more general theories.
\section{Near horizon solution}
Consider, a general theory of gravity and gauge fields with higher derivatives interactions. Let us assume that the theory admits an extremal Reissner-N\"{o}rdstrom(RN) 
black hole 
solution as in (\ref{metric}). Let us consider the general action as,
\bea
S=\int d^4x \sqrt{g}\; {\cal L} (g_{\mu\nu},F_{\mu\nu},f_I),
\eea
where $f_I$'s are some couplings in the theory. To include scalar fields, we replace these couplings by some functions of scalars (but not their derivatives) as $f_I(\phi)$, and add a kinetic term for the scalars in the action. Besides the  modifications to the equations of motion of metric and gauge fields, one also has the equations of motion of 
scalar fields which take the form,
\bea
\label{Klein}
\partial_r\left( a^2b^2\partial_r\phi\right) = \frac{1}{b^2}\frac{\partial W_{eff}}{\partial\phi},
\eea 
where $W_{eff}$ comes from the effective Lagrangian density, 
after inserting the ansatz for the metric and gauge field solutions as background fields in
the action.

\vskip 0.2cm

The near horizon behavior of this equation is important. To a first order approximation, we neglect the backreaction and assume the existence of a black hole with double degenerate horizon.
In this case, one expects to have the following forms for $a(r)$ and $b(r)$ near the horizon:
\bea
&& a(r)  \sim  (1-\frac{r_H}{r}),  \crcr
&& b(r)  \sim  r .
\eea
Now, for a regular solution of $\phi(r)$ near the horizon, the left hand side 
of (\ref{Klein}) vanishes, which in turn implies that $W_{eff}$ should have a 
critical point as $r$ approaches $r_H$. Let us denote this critical point 
by $\phi_{i0}$, such that:  
\bea\label{critical}
\frac{\partial W_{H}}{\partial\phi_i}\mid_{\phi_0}= 0.
\eea

Notice that in contrast to the previous section where $V_{eff}$ was $r$-independent, 
here $W_{eff}$ generically depends on $r$, but its extremum will be meaningful 
only at $r=r_H$. Thus, we use the subscript $H$ which indicates the value of 
$W_{eff}$ at the horizon. 

\vskip 0.2cm

In the other limit at large $r$, if we assume an asymptotically flat background, then the scalar field should appear as a moduli field with arbitrary asymptotic values. In this
way, one can obtain the attractor mechanism, which means that, for any arbitrary values of moduli
fields in the asymptotic region, the scalar fields always approach 
fixed critical values $\phi_{i0}$ at the horizon. 

\vskip 0.2cm

In \cite{Goldstein:2005hq}, this behavior was studied by the perturbative methods 
described above. That is, by setting the asymptotic values of the scalar fields 
equal to their critical values and then examining what happens when the scalars take values at asymptotic infinity which are somewhat different from their attractor values, $\phi_i=\phi_{i0}$.

\vskip 0.2cm

To this end, one starts with a study of the scalar field equation to first order in 
perturbation theory, in the RN geometry without including backreaction. Let $\phi_i$ be a eigen mode of the matrix of second derivatives, $W_{ij}=\partial_i\partial_j W_{eff}$. Then, 
denoting $\delta \phi_i \equiv \phi_i-\phi_{i0}$, 
neglecting the gravitational backreaction and working to a 
first order in $\delta \phi_i$, we find that eq.(\ref{Klein}) takes the form,
\be
\label{eqpertur}
\partial_r\left( (r-r_H)^2\partial_r \delta\phi_i \right) = \frac{1}{r_H^2}w_{2i} \delta \phi_i,
\ee
where $w_{2i}$ is the relevant eigenvalue of $W_{ij}$. In the vicinity of horizon, we have replaced $b(r)$ by $r_H$.  Asymptotically, as $r\rightarrow \infty$, the effects of the gauge fields  and higher derivatives interactions die away and eq.(\ref{eqpertur}) reduces to the equation of motion for a free field in flat space. This has 
two expected solutions, $\delta \phi_i \sim const.$ and $\delta \phi_i \sim 1/r$, both of which are well behaved. One can now notice that the second order differential equation is regular at all points in between the horizon and infinity. So once we choose the non-singular solution in the vicinity of the horizon it can be continued to infinity without blowing up.

\vskip 0.2cm

Now one can easily find the solutions of the linearized equations (\ref{eqpertur}), 
to be \cite{Goldstein:2005hq}:
\bea
\phi_i(r)= C \left( 1-\frac{r_H}{r}\right)^{\lambda_i}
\eea
where,
\bea \label{lambda}
\lambda_i=\frac{1}{2} \left( -1\pm\sqrt{1+\frac{4 w_{2i}}{r_H^2}}\right).
\eea
With a  plus sign, this represents a regular solution at the horizon, provided $w_{2i}$ is 
non-negative. Further, in eqn.(21) $C$ is an arbitrary constant, which 
denotes the asymptotic value of the scalar field.

\vskip 0.2cm

Next, one includes the gravitational backreaction. The first order perturbations in the scalars source a second order change in the metric. The resulting equations for metric perturbations are regular between horizon and infinity. The analysis near the horizon and at the infinity shows that a double-zero horizon black hole solution continues to exist which is asymptotically flat after including the perturbations. In case of \cite{Goldstein:2005hq} the analysis was extended to all orders in perturbation theory analytically and it
was found that the attractor mechanism works to all orders without any extra conditions.

\vskip 0.2cm

To conclude, the key feature that leads to the attractor behavior is the fact that solutions to the linearized equation for $\delta \phi$ are well behaved as $r\rightarrow \infty$, and under some conditions there are regular vanishing  solutions near the horizon. If one of these features fails then the attractor mechanism may not work. For example, adding a mass term to the scalars, results in one of the two solutions at infinity diverging. Now, as discussed in \cite{Goldstein:2005hq}, it is typically not possible to match the well behaved solution near the horizon to the well behaved one at infinity and this makes it impossible to turn on the dilaton perturbation in a non-singular fashion. In the following sections we consider some interesting examples to study the attractor mechanism.

\vskip 0.2cm

Before going through the examples, we wish to  make a comment about the entropy of the black hole. Indeed, following the Wald's formula or using the prescription for the entropy functional given in \cite{Sen:2005wa}, one finds the entropy to be proportional 
to $W_H(\phi_0)$ (see \cite{Sen:2005wa,Sen:2005iz,Prester:2005qs,Alishahiha:2006ke}).   
\section{$R^2$-Gravity}
In this section, we consider a general $R^2$ action coupled to the moduli
field and add it to the Lagrangian (\ref{EHaction}):\footnote{For
simplicity we consider a single moduli field. The generalization
to multi-moduli fields is straightforward. } 
\bea 
\mathcal{L}_2=
G(\phi)\left(\eta R^2+ \beta R_{\mu\nu}R^{\mu\nu}+ \delta
R_{\mu\nu\rho\sigma} R^{\mu\nu\rho\sigma}\right). 
\eea
The most general metric with spherical symmetry is of the form: 
\bea \label{cmetric} 
ds^2= -a(r)^2
dt^2+\frac{1}{c(r)^2a(r)^2}dr^2 +b(r)^2 d\Omega_2^2.
\eea 
Although it is possible to set $c(r)=1$ by a redefinition of the radial
coordinate, we keep $c(r)$ free to derive the complete set of
equations, including the Hamiltonian constraint, from the one
dimensional action. 
\vskip 0.2cm
The gauge field background is same as in eqn. 
(\ref{gaugefield}). Inserting the metric (\ref{cmetric}) in
the action, we find the following one dimensional action\footnote{This is the same as the case of action (\ref{11daction}) where beside the $V_{eff}$ term, all the other terms (being gauge-background independent) are simply derived by inserting the metric ansatz into the original action. $V_{eff}$ term can be reached by inserting the gauge field solution with a sign modification.}: 
\bea
S &=&
\int dr \;\left[\frac{1}{c} - \frac{c}{2} \left(a^2 b^2\right)''-c a^2 b b''
-\frac{V(\phi)}{b^2 c}-2 b a^2 b' c'-b^2 a c' a' \right. \crcr
&& + \frac{1}{b^4}G(\phi) \left\lbrace 2\eta\left(4ac^2a'bb'+cb^2a'c'a+a^2c^2b'^2+c^2b^2aa''+2bca^2b'c' -1 \right. \right. \crcr
&&\left. +c^2b^2a'^2 +2bc^2a^2b'' \right)^2 + \beta \left( 1-2 a^2 c^2 b'^2+c^4 b^4 a^2 a''^2+3 c^4 b^2 b''^2 a^4-2 b c a^2 b' c' \right.  \crcr
&&+\;8 c^4 b^2 b'' a^3 a' b'+\;8 c^3 b^2 b'^2 c' a^3 a'+\;2 b c^4 a^4 b'' b'^2+\;4 a^3 c^4 a' b b'^3+\;3 c^2 b^2 b'^2 c'^2 a^4   \crcr
&&+2 c^4 b^3 a^3 a'' b''+\;2 c^3 b^3 a^3 a'' b' c'+\;2 c^3 b^3 a' c' a^3 b''+2 c^2 b^3 a' c'^2 a^3 b'+2 c^4 b^3 a'^2 b'' a^2  \crcr
&&+\;6 c^3 b^2 b'' a^4 b' c'+\;2 c^3 b^4 a^2 a'' a' c'+\;2 c^4 b^4 a a'' a'^2+4 c^4 b^3 a^2 a'' a' b'+\;c^2 b^4 a'^2 c'^2 a^2  \crcr
&&+2 c^3 b^4 a'^3 c' a+6 c^3 b^3 a'^2 c' a^2 b'+4 c^4 b^3 a'^3 a b'+8 c^4 b^2 a^2 a'^2 b'^2+c^4 b^4 a'^4 +\;a^4 c^4 b'^4   \crcr
&&\left. -2 b c^2 a^2 b''-\;4 a c^2 a' b b'+\;2 a^4 c^3 b'^3 b c'\right)  
+\;2 \delta \left( 1+\;2 c^3 b^4 a^2 a'' a' c'+\;2 c^4 b^2 b''^2 a^4 \right. \crcr
&&+c^4 b^4 a^2 a''^2+4 c^3 b^2 b'' a^4 b' c'+c^2 b^4 a'^2 c'^2 a^2 
+4 c^4 b^2 b'' a^3 a' b'  -2 a^2 c^2 b'^2+c^4 b^4 a'^4\crcr 
&&+\;\;2 c^2 b^2 b'^2 c'^2 a^4+\;4 c^3 b^2 b'^2 c' a^3 a' +\;2 c^4 b^4 a a'' a'^2+\;2 c^3 b^4 a'^3 c' a +\;4 c^4 b^2 a^2 a'^2 b'^2 \crcr
&&\left. \left.\left.  +a^4 c^4 b'^4 \right) \right\rbrace \right] 
\eea
One derives the equations of motion by varying the above action
with respect to $a$, $b$, $c$ and $\phi$. Then we can put
$c(r)=1$, and the equation for $c$ turns out to be the Hamiltonian
constraint. We call these equations as $EqA$, $EqB$, $EqC$ and $Eq\Phi$, 
respectively\footnote{Here we avoid writing these
equations which are very lengthy. We have used the \emph{maple}
package to derive them.}. Though the total effective potential $W_{eff}$ introduced in the previous section has a complicated form in this theory, it is not too hard to find its critical point at the horizon as we show below. 
\vskip 0.2cm
It is well known that these equations
admit $AdS_2 \times S^2$ as a solution with constant moduli.
However we want to address the attractor behavior in solutions
which are asymptotically flat. In view of the fact that the four
equations governing $(a(r),b(r),\phi(r))$ are a set of highly
complicated coupled differential equations of order four. To solve
these equations, we follow the Frobenius method. As a variable of expansion
we define $x \equiv 1-\frac{r_H}{r}$, ranging from 0 to 1 to cover
$r\geq r_H$ completely. 
\vskip 0.2cm
In fact, requiring that the solution:
\begin{enumerate}
\item
be extremal:  meaning that $a(r)=(r-r_H)^2H(r)$, with $H(r)$
being analytic at the horizon, $r=r_H$.
\item be
asymptotically flat: meaning that $(a(r),b(r),\phi(r))$ tend to
$(1,r,\phi_\infty)$ at asymptotic infinity.
\item be regular at the horizon,
\end{enumerate}
the most general Frobenius expansions of
$a(r)$,$\;b(r)$ and $\phi(r)$ take the form:
\bea\label{aexpan}
a(r)&=&\;(x+x^{\lambda_1}\;\sum_{n=2}^{\infty}a_n x^n),  \\
\label{bexpan}
b(r)&=&\; \frac{r_H}{(1-x)}\;(1+x^{\lambda_2}\sum_{n=1}^{\infty}
b_nx^n),  \\
\label{phiexpan}
\phi(r)&=&\; (\phi_0+x^{\lambda_3}\;\sum_{n=1}^{\infty}\phi_n x^n),
\eea 
with $\lambda_i \geq 0$. 
\vskip 0.2cm
For concreteness, from now on
we set both the functions $G(\phi)$, which is the coupling of moduli
field to $R^2$, and $V(\phi)$, which is the $U(1)$-background
contribution to the effective potential, to be of standard
dilatonic forms, as \emph{inspired by string theory}. More
precisely, denoting the attractor value of moduli with $\phi_0$ we set,\footnote{Notice that with this parametrization, $G(\phi_0)=g_0$ and $V(\phi_0)=v_0$.} 
\bea\label{VG}
G(\phi)\equiv g_0\;e^{\sigma(\phi-\phi_0)}\;\;\;\;;\;\;\;V(\phi) \equiv
v_0\;e^{\mu(\phi-\phi_0)}.
\eea

We mention that although, in comparison to (\ref{eff}), $V(\phi)$ of (\ref{VG}) is of pure magnetic (electric) type, the case given in (\ref{VG}) is rich enough for our study since the new function $G(\phi)$, mathematically, compensates the role played by the lacking electric (magnetic) term.  

\begin{center}
\underline{{\bf Zeroth order results}}
\end{center}
For the case at hand, it is consistent to set $\lambda_i=0$. $EqA$
is automatically satisfied and either of $EqB$ or $EqC$ implies,
\bea\label{v0} 
v_0=r_H^2 , 
\eea 
while $Eq\Phi$ gives:
\bea\label{v1}
G(\phi_0)=\frac{r_H^2\;\mu}{2\;\sigma\;(\beta+2\delta)}\;.
\eea
Notice that equations (\ref{v0}) and (\ref{v1}), together, determine both of the attractor value of the moduli filed and the horizon radius in terms of the parameters of the action. In fact both the above results are meaningful to us. Due to (\ref{v0}) the Bekenstein-Hawking entropy of the solution is given by the value of the parameter $v_0$, up to a numerical prefactor. The
information encoded in (\ref{v1}) becomes clearer when we note
that this relation, together with $(\ref{VG})$, implies:
\bea\label{atra} 
\frac{dV}{d\phi}\;-\;2(\beta+2\delta)\frac{dG}{d\phi}\;=0\;,
\eea
which is exactly (\ref{critical}) with,
\bea\label{Weff}
W_H(\phi) \equiv V(\phi)-2(\beta+2\delta) G(\phi).
\eea
This in fact fixes $\phi_0 $ at its extremum point.
From (\ref{phiexpan}), $\phi_0 = \phi(r_H)$ and so the value of
the moduli field is fixed at the horizon, regardless of any other
information. Thus to complete the
proof of the attractor behavior, we should be able to show that the four sets of
equations of motion, denoting a coupled system of differential
equations, admit the expansions (\ref{aexpan}), (\ref{bexpan}) and
(\ref{phiexpan}). Furthermore, one should see that there are 
solutions to all orders in the $x$-expansion with arbitrary 
asymptotic values at infinity, while the value at the horizon is fixed to be $\phi_0$. 
The existence of a complete set of solutions with desired boundary conditions (considering the 
fact that we have coupled non-linear differential equations) by itself is not trivial. Moreover, 
it is easy to show that, in our theory, there is no asymptotically flat solution 
with everywhere constant moduli. 

\begin{figure}
 \begin{center}
  \includegraphics[scale=.6]{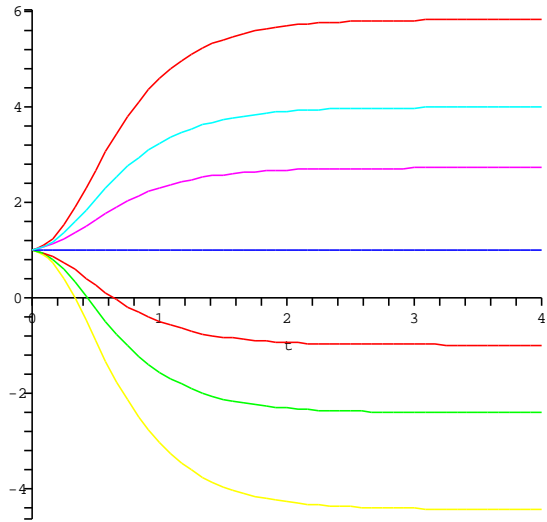} 
 \end{center}
\caption{$\phi(r)$ vs. $\log(r)$, where numerical coefficients are $\eta=1, \beta=2, \delta=3, \mu=5$ and $g_0=4$. Different curves represent different asymptotic values for $K=\phi_{\infty}$. The attractor point is $\phi_0=1$ at the horizon,  $r_H=1$.}
\end{figure}

\begin{center}
\underline{{\bf Higher order results}}
\end{center}
We solve the system of equations $(EqA,EqB,Eq\Phi,EqC)$ order by order in the $x$-expansion. 
To first order, we find that one variables, say $\phi_1$, can not be fixed by 
the equations. Let us denote the value of $\phi_1$ as $K$. We thus find 
$a_2$ and $b_1$ as a function of $K$. 
One can check that at any order $n>2$, one can substitute the
resulting values of $(a_{m},b_{m-1},c_{m-1})$, for all $m \leq n$
from the previous orders.  Then $(EqB,Eq\Phi,EqC)$ of the
current order together with $EqA$ of order $(n-1)$,
\emph{consistently} give,
\bea\label{resultn}
b_n=b_n(K,g_0)\;\;\;;\;\;\;a_{n+1}=a_{n+1}(K,g_0)\;\;\;;\;\;\;\phi_{n}=\phi_{n}(K,g_0)\;\;\;.
\eea 
as polynomials of order $n$ in terms of $K$.
\vskip 0.2cm
It is worth noting that $K$ remains a free parameter to all
orders in the $x$-expansion. From (\ref{aexpan}), (\ref{bexpan}) and
(\ref{phiexpan}), the asymptotic values of $(a(r),b(r),\phi(r))$
are given by a sum of all the coefficients in the x-expansion of
the corresponding function. Owing to the result
(\ref{resultn}) we observe that $(a_\infty,b_\infty,\phi_\infty)$
are free to take different values, given different choices for $K$
\footnote{We do not address the convergence of the series in
detail, but it would be the case for small enough values for $|K|$.}.
\vskip 0.2cm
The arbitrary value of  $\phi$ at infinity is $\phi =\phi_\infty$, while its value at the horizon is fixed to be $\phi_0$. This is the very attractor mechanism which we were looking for in this model. Figure 1 shows $\phi$ vs. $r$ in logarithmic scale with different asymptotic values $K$. 
\section{Black Holes in Gauss-Bonnet Gravity}   
Since the Gauss-Bonnet case is a special sector of general $R^2$
gravity, it is obvious that any solution of the form considered in the previous section, enjoys the attractor
mechanism. Now we relax the analyticity condition and, instead, only
require the solutions to be regular at the horizon. This
relaxation, enables us to cover a class of solutions connected
to $G\equiv0$ case discussed in \cite{Goldstein:2005hq}, which are
not analytic at $x=0$.
\vskip 0.2cm
Black holes in higher order gravity are being actively pursued. It
is important to note that string theory predicts the
Einstein-Hilbert action to be modified by higher order curvature
corrections. The leading correction is quadratic in curvature.
Typically, only certain special corrections are considered. At
quadratic order, the. combination which is considered is the
Gauss-Bonnet term:
\bea 
\mathcal R_{\gamma\delta\lambda\sigma}
  R^{\gamma\delta\lambda\sigma}-4 R_{\gamma\delta}R^{\gamma\delta}
  +R^2
\eea 
We are interested in an Einstein-Gauss-Bonnet theory coupled to the
gauge fields and scalars. Since the Gauss-Bonnet term in four
dimensions is a topological term, the coupling to the scalar fields
is crucial. Without this coupling, the Gauss-Bonnet term does not
contribute to the equations of motion. In that case, we are
left with the RN solution reviewed in section 2.
\subsection{Equations of motion and attractors}
Consider a gravity theory coupled to $U(1)$ gauge fields and scalars in 
the presence of higher order corrections of the Gauss-Bonnet type. 
The system is dictated by following bosonic action:
\bea
\label{actiongb}
  S=\frac{1}{\kappa^{2}}\int d^{4}x\sqrt{-g}\bigg{(}\;R-2(\partial\phi_i)^{2}-
  f_{ab}(\phi_i)F^a_{\mu \nu} F^{b \ \mu \nu}   - \frac{1}{2} {\tilde f}_{ab}(\phi_i) F^a_{\mu \nu}
  F^b_{\rho \sigma} \epsilon^{\mu \nu \rho \sigma} &&\cr\cr 
+ \alpha G(\phi)\mathcal (R_{\gamma\delta\lambda\sigma}
  R^{\gamma\delta\lambda\sigma}-4 R_{\gamma\delta}R^{\gamma\delta}
  +R^2)\bigg{)}, &&
\eea
where $\alpha$ is the Gauss-Bonnet coefficient with dimensions of (length)$^2$ and is positive in Heterotic string theory. The function $G(\phi)$ is arbitrary and signifies a dilaton-like coupling.  
\vskip 0.2cm
Let us also assume that all quantities are functions of $r$ and take
the following ansatz for the metric and gauge fields of a Reissner-N\"{o}rdstrom-
Gauss-Bonnet (RNGB) black hole:
\bea
  ds^{2} & = & -a(r)^{2}dt^{2}+a(r)^{-2}dr^{2}+
b(r)^{2}d\Omega^{2},\label{metric2}\\
  F^a & = &  f^{ab}(\phi_i)(Q_{eb}-{\tilde f}_{bc}Q^c_m)
\frac{1}{b^2} dt\wedge dr + Q_m^a sin \theta  d\theta \wedge d\phi,
\eea
Plugging\footnote{See footnote 7.} the above ansatz in the action one finds the following one-dimensional action:
\bea \label{12daction}
  S = \frac{1}{\kappa^{2}}\int dr\left[2-(a^{2}b^{2})^{''}-2a^{2}bb^{''}
    -2a^{2}b^{2}(\partial_{r}\phi)^{2}
-\frac{2 V_{eff}(\phi_i)}{b^{2}} \right. &&\cr\cr
+ \left. 8\,\alpha\,G(\phi)  \partial_r\left(aa'(-1 + a^2b'^2)\right)
 \right].&&
\eea
The Hamiltonian constraint, which
must be imposed in addition takes the form:
\bea \label{constraint}
 -1 + a^{2}b^{'2}+\frac{a^{2'}b^{2'}}{2}  +\frac{1}{b^{2}} V_{eff}(\phi_i) - a^{2}b^{2}\phi'^2 =-4\alpha G' \left( aa'-3a^3a'b^{'2}
\right), 
\eea
where a primes indicates derivation with respect to $r$. 
\vskip 0.2cm
The equations of motion following from (\ref{12daction}) are\footnote{It is easily checked that the RHS of the first equation, if evaluated at the horizon, 
leads to the effective potential $W_{eff}=V + 4\; \alpha \; G$. This concides with the general effective potential (\ref{Weff}) for the G.B. parameters $(\beta,\delta)=(-4,1)$, if we re-absorb $\alpha$ into the function $G$.}:
\bea
\left(a^2 b^2\phi'\right)' = \frac{1}{2b^2}\frac{d V_{eff}}{d\phi} &+& 2\a\frac{dG}{d\phi}\left(aa'\left(1-a^2b'^2\right)\right)', \\
\left( a^2b\right)''-a'^2b -2aa'b' + a^2b\phi'^2+\frac{V(\phi)}{b^3} &=& 4\a\left( \left( a^3a'b'G'\right) '+a^3a'b'G'' \right),  \\
bb''+b^2 \phi'^2 &=& \a \left( -2 G'' +2 a^2 \left( G'b'^2\right)'  \right).
\eea
To solve the above equations, we take two different approaches: $x$-expansion 
and $\a$-expansion as explained in the following subsections. 

\subsection{$x$-Expansion series solutions}
By $x$-expansion, we mean a set of solutions as a Frobenius series given in 
(\ref{aexpan})-(\ref{phiexpan}). We take a 
common $\lambda_i\equiv \lambda$ for all the solutions and write the series as: 
\bea
\phi(r)&=& \phi_0+K x^\lambda +\cdots,   \cr\cr
a(r)&=& x + a_1 x^{\lambda+1}+\cdots,  \cr\cr
b(r)&=& r (1+ b_1 x^\lambda+\cdots ). \nonumber
\eea
Let us also consider Taylor series expansions for $V(\phi)$ and $G(\phi)$ as follows,
\bea \label{Vtaylor}
V(\phi)&=&v_0 + v_1 (\phi-\phi_0)+ \frac{1}{2}v_2 (\phi-\phi_0)^2 +\cdots ,\\
\label{Gtaylor}
G(\phi)&=&g_0 + g_1 (\phi-\phi_0)+ \frac{1}{2}g_2 (\phi-\phi_0)^2 +\cdots \:.
\eea
Since integer values of $\lambda$ are included in the results of 
section 4, we look for non-integer values $\lambda > 0$, which give 
us a regular (though non-analytic) set of solutions at the horizon. 
\vskip 0.2cm
By a careful investigation near the horizon, for the lowest power of $x$ which is $x^\lambda$, one can solve the set of equations together and find the non-trivial solutions as:
\bea \label{lambdaGB}
\lambda&=&\frac{1}{2}\left(-1+\sqrt{1+2\frac{(v_2+4g_2)}{r_H^2}}\right) , 
\eea
\bea  \label{attractor}
v_0=r_H^2, \;\;\;\;\;\; v_1=-4g_1,
\eea
with $g_1=a_1=b_1=0$. However, $v_2$, $g_2$ and $K$ are undetermined to this order. The value of $\lambda$ in (\ref{lambdaGB}) is precisely what we already found in eqn. (\ref{lambda}) with 
$W_H=V(\phi)+4G(\phi)$. The second equation in (\ref{attractor}) is the extremum condition for $W_H$ which gives the attractor value $\phi_0$ at the horizon. Notice that we 
are faced with an extra condition $g_1=0$, which indicates that $G$, $V$ and $W_H$ are at their 
extremum at the horizon, simultaneously. Such a case would be the only situation where a non-integer $\lambda$ can be found. Otherwise we have to choose $\lambda=0$ for $g_1 \neq 0$. \vskip 0.2cm
The regularity condition for $\phi$ indicates that $\lambda$ should be non-negative and it in turn gives $v_2+4g_2 \ge 0$. This again means that $W_H$ is minimum at its extremum point $\phi_0$.
\vskip 0.2cm
Higher order terms can be derived in a similar fashion. The important point is that, 
due to the non-linear nature of equations, they are a mixture of different powers of $\lambda$, like $x^{n\lambda}$ as well as $x^{n\lambda+m}$. To order these powers, we assume $0<\lambda <1$. Then the next leading term would be $x^{2\lambda}$. For higher order terms, since $\lambda$ is already known from the first order result (\ref{lambdaGB}), we can distinguish the next order from $x^{3\lambda}$ and $x^{\lambda+1}$. For small enough $\lambda$, it shows that we are generating a power series, $x^{n\lambda}$ as 
argued in \cite{Goldstein:2005hq}. 
\vskip 0.2cm
Notice that in contrast to the analysis of section 3, here we considered all the equations simultaneously. This first means that, in principal, we are taking the backreaction into account. Secondly, since we are dealing with a higher derivative theory, besides the so-called Klein-Gordon equation for $\phi$ field. Other equations also involve the second derivative of $\phi$ and are important in the dynamics of $\phi$. So, 
they should be investigated as well. 

\subsection{$\a$-Expansion}
Motivated by low energy effective actions of string theory, it is be reasonable to consider $\a$ as a small parameter (proportional to $\a'$) and try to solve the set of equations perturbatively 
in the $\a$ parameter. Obviously, to zeroth order, we start with the results in
\cite{Goldstein:2005hq}. Let us consider   
\bea
\phi(r)&=& \phi^{(0)}+\a \phi^{(1)}(r) + \a^2\phi^{(2)}(r)+\cdots,   \cr\cr
a(r)&=& \left(1-\frac{r_H}{r}\right) + \a a^{(1)}(r)+\a^2 a^{(2)}(r)+\cdots,  \cr\cr
b(r)&=& r \left(1+ \a b^{(1)}(r) +\a^2 b^{(2)}(r)+\cdots \right), \nonumber
\eea
together with Taylor expansions  (\ref{Vtaylor}) and (\ref{Gtaylor}). 
\vskip 0.2cm
Take $\phi^{(0)}=\phi_0$, the constant solution at zeroth order. Then we find,
\bea
\phi^{(1)}=\left( 1-\frac{r_H}{r}\right) ^{(-1+\sqrt{1+2v_2/r_H})/2} +P_4(r),
\eea
where $P_4$ is a function of $r$ which behaves like a polynomial $x$ of order 4 near the horizon. Moreover $a^{(1)}(r)$ and $b^{(1)}(r)$ can be set to zero consistently. It is worth mentioning that the attractor equation (\ref{attractor}) should remain valid to all orders in 
$\a$, so that when expanding in powers of $\a$:
\bea
\frac{d W_H}{d\phi}&=& \frac{d}{d\phi} \left( V(\phi)+4\a G(\phi)\right)|_H  \cr\cr
&=&v_1+\left( v_2 \phi^{(1)}_H+4g_1\right) \a+\frac{1}{2}\left( v_2\phi^{(2)}_H+v_3(\phi^{(1)}_H)^2+8g_2\phi^{(1)}_H\right) \a^2+{\cal O}(\a^3), \cr\cr 
&=&0.
\eea
Solving at each order in $\a$, one finds:
\bea
v_1&=&0, \cr\cr
\phi^{(1)}_H&=&-\frac{4g_1}{v_2},  \cr\cr
\phi^{(2)}_H&=&\frac{16}{v_2^3} \left( 2g_1g_2v_2 -g_1^2v_3\right), \nonumber
\eea
where the first equation is the extremum condition to zeroth order in $\a$ 
and the other two equations give the correct boundary value of the first and 
second order fields at the horizon, respectively. 
\begin{figure}
 \begin{center}
  \includegraphics[scale=.6]{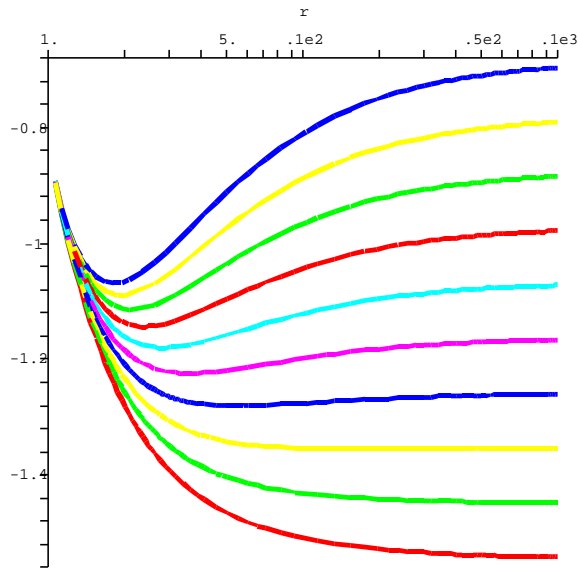} 
 \end{center}
\caption{$\phi(r)$ vs. $\log(r)$, where numerical coefficients for the potentials are $g_2=v_{i\ge2}=1$ and $g_1=2$. Different curves represent different asymptotic values for $\phi_{\infty}$. The attractor point is $\phi_0=0$ at the horizon,  $r_H=1$.}
\end{figure}

At the second order, we one can find non-zero $a^{(2)}(r)$, $b^{(2)}(r)$ and $\phi^{(2)}(r)$, in terms of some integration constants. These constants are subject to 
boundary conditions which we already mentioned in section 4. Of course we have 
enough integration constants to satisfy all the boundary conditions. Especially, 
for the scalar field, we set its value at the horizon to be $\phi_0$ and at infinity 
it takes any free value. This proves the attractor behavior of the system for 
scalar field moduli. 
\vskip 0.2cm
The $x$-expansion of $\phi^{(2)}(r)$ shows that it includes terms like $x^n\log{x}$ which indicate the non-analytic behavior, though regular, near the horizon. 
To avoid quoting lengthy results, we demonstrate our results in figure 2. 

\section{Conclusion}

In this paper, we studied the attractor mechanism in a theory of gravity coupled
to gauge fields and scalar fields, with higher derivative $R^2$ terms in the action.
By investigating solutions of the equations of motion, we observed the attractor behavior explicitly. We looked for all possible solutions which admit the criteria of being regular at the horizon and free in the asymptotic region. 
\vskip 0.2cm
The near horizon analysis done in
section 3 (see also \cite{Sen:2005wa}), shows the criteria for attractor behavior. 
Although, the existence of a consistent set of solutions involving the backreaction 
remains far from obvious. 
In $R^2$ gravity and especially in Gauss-Bonnet theory, we solved the equations 
for the background and scalar fields simultaneously which means inclusion of 
backreaction effects. 
\vskip 0.2cm
For the Gauss-Bonnet theory, given flat asymptotic boundary condition for backgrounds, the regularity of scalar fields at the horizon is a sufficient (and obviously necessary) condition to meet the attractor mechanism. The regularity at the horizon, in the non-analytic sector, sets some restriction on the effective potential $W_H$ which is indeed its minimality condition at its critical point. 
\vskip 0.2cm
The solution with analytic behavior at the horizon appears as a new
branch in the set of attractor solutions which gives no minimality restriction on the effective potential, $W_H$. So we observe that, compared to the case studied in \cite{Goldstein:2005hq}, turning on the $R^2$ terms and coupling them to the moduli in the form of (\ref{VG}), a new sector of solutions appear for which the extremality of the effective potential is enough to produce the attractor behavior.
\vskip 0.2cm
Although, for the analysis in this paper we considered asymptotically flat 
space times, with the same technology, it is easy to check that attractor mechanism works as well for asymptotically (A)ds backgrounds.
\begin{center}
{\large {\bf Acknowledgement}}
\end{center}
Authors would like to thank B. Acharya, M. Alishahiha, A. Dabholkar, R. P. 
Jena, K.S. Narain, S. Randjbar-Daemi and M.M. Sheikh Jabbari 
for useful discussions and  comments, and 
K. Goldstein for informing  us of \cite{Gibbons:1996af}. 
A.T. also thanks the Department of Theoretical Physics at TIFR for 
friendly hospitality. H. Y would like to thank the High Energy Group of 
the Abdus Salam International Centre for Theoretical Physics for hospitality and support.

\end{document}